\documentclass[manuscript,screen,review]{acmart}
\usepackage{subfigure}

\AtBeginDocument{%
  \providecommand\BibTeX{{%
    \normalfont B\kern-0.5em{\scshape i\kern-0.25em b}\kern-0.8em\TeX}}}

\acmConference[AIMLSystems '21]{AIMLSystems '21}{October 21--23, 2021}{Bangalore, India} 

\newcommand{\verlpy}{VeRLPy }

\begin{document}
\title{VeRLPy: Python Library for Verification of Digital Designs with Reinforcement Learning}

\author{Aebel Joe Shibu}
\email{aebeljs@gmail.com}
\affiliation{%
  \institution{Indian Institute of Technology Madras}
  \country{India}
}
\authornote{Both authors contributed equally to the research}

\author{Sadhana S}
\email{sadhashan118@gmail.com}
\affiliation{%
  \institution{Indian Institute of Technology Madras}
  \country{India}
}
\authornotemark[1]

\author{Shilpa N}
\email{shilpa16b039@gmail.com}
\affiliation{%
  \institution{Indian Institute of Technology Madras}
  \country{India}
}

\author{Pratyush Kumar}
\email{pratyushkpanda@gmail.com}
\affiliation{%
  \institution{Indian Institute of Technology Madras}
  \country{India}
}




\begin{abstract}
Digital hardware is verified by comparing its behavior against a reference model on a range of randomly generated input signals. 
The random generation of the inputs hopes to achieve sufficient \textit{coverage} of the different parts of the design. 
However, such coverage is often difficult to achieve, amounting to large verification efforts and delays. 
An alternative is to use Reinforcement Learning (RL) to generate the inputs by learning to prioritize those inputs which can more efficiently explore the design under test.
In this work, we present VeRLPy an open-source library to allow RL-driven verification with limited additional engineering overhead. 
This contributes to two broad movements within the EDA community of (a) moving to open-source toolchains and (b) reducing barriers for development with Python support.
We also demonstrate the use of VeRLPy for a few designs and establish its value over randomly generated input signals.

\end{abstract}


\maketitle

\section{Introduction}
\label{sec:intro}

Hardware verification is involved after each stage of the Electronic Design Automation (EDA) flow to ensure functional correctness. It is the most expensive and time-consuming part and takes more than 70\% of the development cycle. The typical approach towards such verification is to compare the performance of a device-under-test (DUT) with a reference model on a series of randomly generated test cases. This tends to take a very large number of input test cases to achieve acceptably high coverage of the design. The alternative is to rely on domain expertise to generated test cases reflecting certain constraints. However, this approach is usually more expensive and does not scale to very complex designs. A good survey of these trade-offs is provided in \cite{mledasurvey}. Coverage metrics such as functional, code, and Finite State Machine (FSM) path coverage are used to measure the quality of verification. But significant engineering effort is required in analyzing the results of these coverage outputs and updating the test plans and input stimulus to the DUT to improve the coverage. 

There has been a growing trend in the use of learning-based methods in optimization problems encountered in systems engineering. 
Specifically, the use of Reinforcement Learning (RL) has enabled the optimization of many stateful systems. 
For instance, RL has seen use in resource management \cite{DeepRM, ye2018new, liu2017hierarchical}, network routing \cite{Q-routing, SDN} and most recently in chip placement \cite{mirhoseini2020chip}.
More broadly, RL has achieved dramatic breakthroughs in various domains especially in combination with deep learning (DL). The simplicity of the Markov Decision Process (MDP) framework and the approximation power of neural networks have allowed RL to solve these stateful and complex optimization problems.

Digital verification, and in particular, optimizing the set of input signals to be given to a DUT, can be framed as an RL optimization problem.
This is because of the underlying similarities between an RL environment and the FSM of the DUT. The hardware implementation of the FSM consists of just two types of components, namely, the combinational elements and the memory elements. The contents of the memory elements completely specify the state of the DUT. This inherent stateful nature is precisely what an MDP can model. Further, the optimization objectives of high coverage are implicit and best expressed as reward signals for an RL agent. Thus, the DUT becomes the RL environment, and the RL agent replaces the human intelligence that steers the verification procedure towards higher coverage by guided input stimulus selection. 

This paper is not the first to propose to use RL for digital verification. 
In \cite{better}, the design of a processor subsystem was verified with an RL algorithm to guide input generation.
RL has recently been used more widely in other EDA algorithms such as placement and routing \cite{mirhoseini2020chip, goldie2020placement}. 
Our goal in this paper is to build the first known open-source implementation of a verification framework that has a few properties. 
First, the entire framework is open-source building on other open-source efforts such as cocotb \cite{cocotb} for verification, Verilog for hardware description, and OpenAI Gym \cite{Gym} for the RL framework and base algorithms. 
Second, the entire framework is based on Python, thereby reducing the barriers of entry for hardware and software engineers in further extending the library. 
Third, the framework is designed to build a clear separation of concerns wherein the RL algorithms are well decoupled from the DUT's interface to the verification engine through cocotb. 
This ensures a modular approach wherein an RL algorithm can be replaced without requiring any change at all in the DUT or its interface with the cocotb. 
Finally, the framework already comes with defaults for various components: Verilog-based design, logging of events through cocotb, a range of RL algorithms, and support for extensive logging and visualization.

Apart from describing the design of VeRLPy and how it can be used, we also describe the use of VeRLPy for a couple of designs. 
In particular, we study two designs - a Run Length Encoder (RLE) based compressor which is a component in an open-source accelerator for deep learning that our team is building \cite{shaktimaan}, and an AXI controller as part of an open-source family of processors being designed at IIT Madras \cite{shakti}. 
For both these designs, we show how VeRLPy helps to reduce the number of inputs required for the verification to reach rare combinations of internal state that is essential in discovering design bugs.

Our framework VeRLPy is open-sourced and available at \cite{VeRLPy}. 
We hope that VeRLPy is useful in accelerating research in the intersection of Reinforcement Learning and design verification. 
We also hope that VeRLPy is helpful in designing various RL Gym environments, thereby diversifying to a new class of environments to test RL algorithms on.

The rest of the paper is organized as follows. Section \ref{sec:background} goes over the main building blocks of the VeRLPy framework. Section \ref{sec:framework} explains the architecture and control flow of the framework. Section \ref{sec:experiments} discusses the results from the experiments using the framework, and Section \ref{sec:conclusion} concludes with the possible next steps.

\section{Background}
\label{sec:background}


\subsection{The CocoTB framework for Verification}

Although testbenches for hardware verification have been traditionally written using HDLs, there has been a recent shift towards using cocotb. 
CocoTB (Coroutine-based co-simulation Testbench)\cite{cocotb} is an open-source package using which the test benches for verification can be written in a higher-level general-purpose programming language like Python. 
It needs a simulator to simulate the HDL design and in general, it supports any simulator with industry-standard VPI, VHPI, or FLI interfaces. CocoTB has Driver and Monitor classes which are used for driving inputs and monitoring inputs and outputs. The input drivers are used to drive inputs to DUT. If any input is driven to the DUT, the input monitor coroutine calls back the model function whose output will be given to the scoreboard. Similar to the input monitor, there is an output monitor coroutine that receives output from the DUT and passes the output to the scoreboard class. The scoreboard compares the model output and the output from the DUT. CocoTB majorly interacts only with the periphery of the design and hence it is enough to add only input and output signals from the modules to the respective drivers and monitors. Additional coroutines can also be added to monitor register states by using the hierarchical path of the elements. 

The advantage of being able to use Python for the tests is twofold. Firstly, testbenches are easier to set up since Python is a programming language that is much more accessible and widely used compared to conventional HDLs. Secondly, the testbench logic being completely written in Python provides access to a plethora of open-source software packages which are compatible with Python. This is paramount in letting us efficiently leverage the existing work in the field of AI since a major chunk of the research and development done in this regard has been using Python.

\subsection{Reinforcement Learning and the Gym environment}
Reinforcement Learning (RL) is a sub-field of Machine Learning where an agent learns by interacting with an environment. 
Basic reinforcement is modeled as a Markov Decision Process (MDP) where the environment has a set of states with an unknown transition probability, that the agent interacts with through a defined set of actions and in turn receives a reward signal. 
The agent then combines exploration and exploitation to optimize the actions performed to maximize the cumulative reward.
Each RL episode could be composed of multiple timesteps. In each timestep $t$, the action $a_t$ chosen by the RL agent results in consequences in the RL environment. At the end of each timestep, the agent receives from the environment the updated state $s_t$ and a reward $R_t$ indicating whether the consequences were favorable or not. The agent explores different actions, and the corresponding rewards obtained are used to learn a suitable policy to solve the task.

The development of OpenAI Gym \cite{Gym} led to the definition of a specific structure that RL environments could adhere to, thus filling the need for proper standardization of RL environments across diverse domains. Since Gym does not make any assumptions about the structure of the RL agent, it is compatible with the various numerical computation and RL algorithm packages available. This decoupling has allowed Gym to be used to develop environments for Go \cite{GymGo}, robotic simulations \cite{PyBullet}, financial trading \cite{GymTrading}, neural architecture search \cite{NASGym}, and in VeRLPy for hardware verification.

A variety of RL algorithms have been developed over the years. The broad classes of these algorithms include value function based methods like Q-learning \cite{watkins1992q} and Deep Q-Networks (DQN) \cite{mnih2015human}, policy search methods like Trust Region Policy Optimization (TRPO) \cite{schulman2017trust} and Proximal Policy Optimization (PPO) \cite{schulman2017proximal}, and actor critic methods like Soft Actor Critic (SAC) \cite{haarnoja2018soft}. The study of these algorithms is an active area of research. In this paper, we use these algorithms as a replaceable module as enabled by the Gym framework. Specifically, for the experiments reported later, we use the implementation of SAC provided by Stable Baselines3 \cite{sb3}.

\section{The \protect \verlpy Framework}
\label{sec:framework}

\subsection{Software Architecture}

\begin{figure}[htbp]
\subfigure[A top level view of the framework 
\label{layer_block}]
{\includegraphics[width=4cm, height=3.5cm]{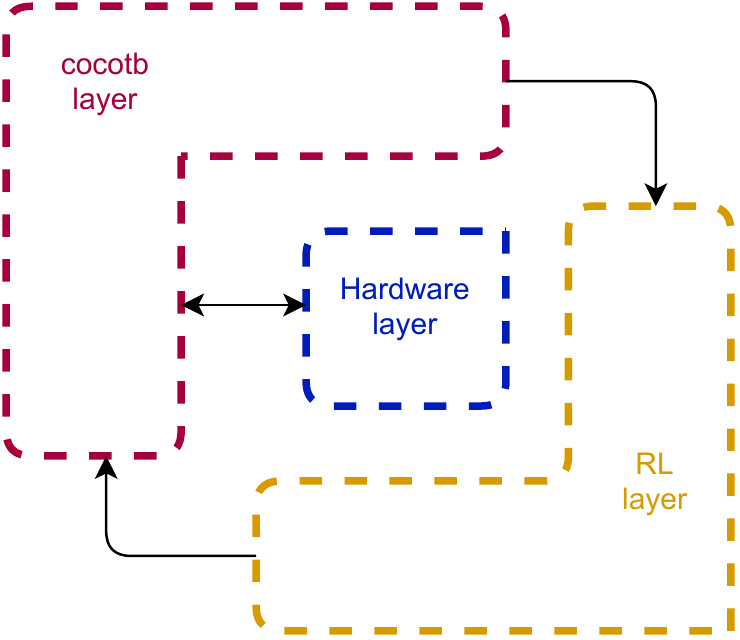}}
\subfigure[Block diagram of the complete framework with the component layers demarcated using magenta (cocotb layer), blue (Hardware layer) and orange (RL layer) \label{RL_verif_without_notation}]{\includegraphics[width=9cm,height=7cm]{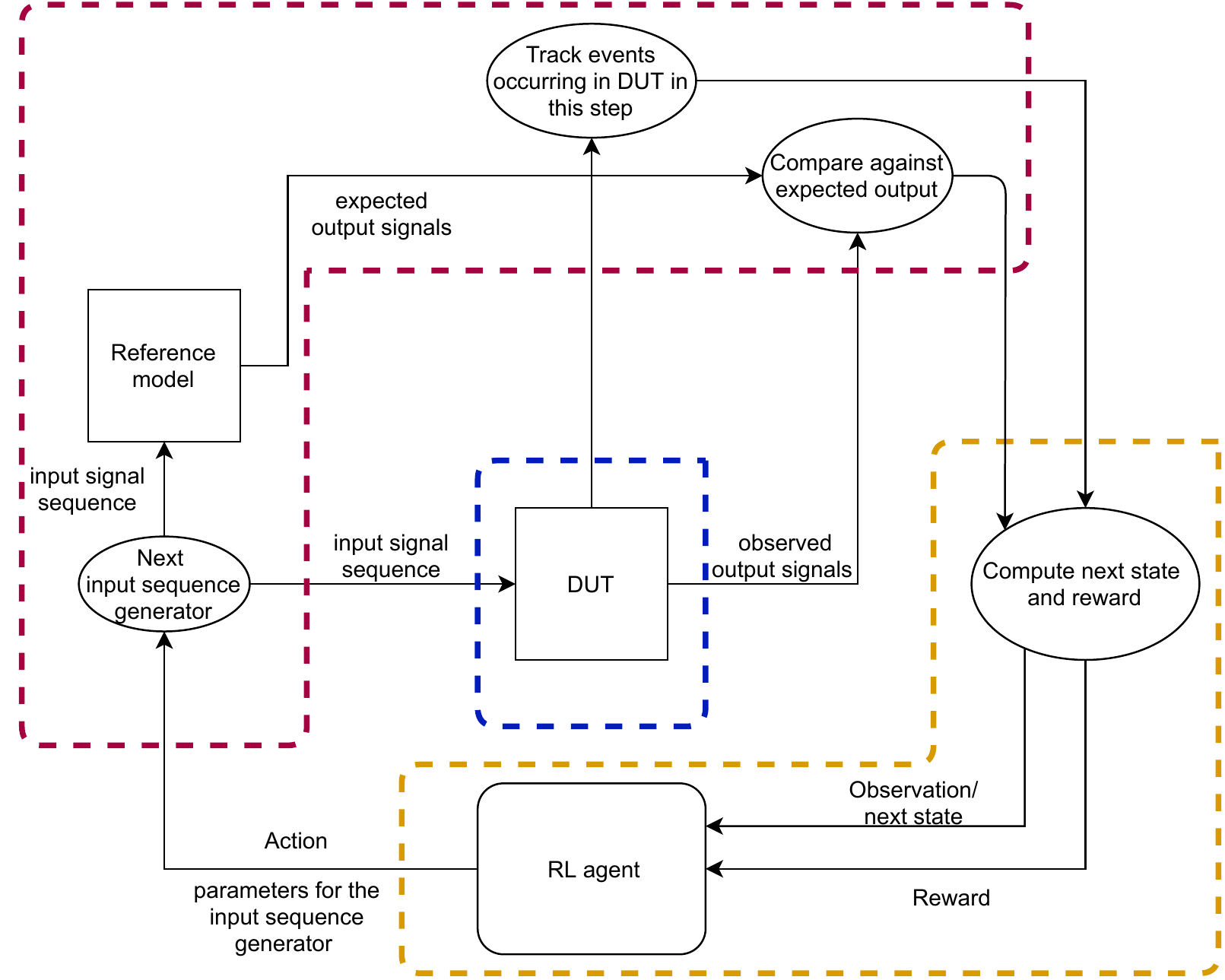}}
\caption{The Software Architecture of VeRLPy. }
\end{figure}

The software architecture of the VeRLPy framework consists of three layers that interact with one another, as shown in Fig. \ref{layer_block}. The hardware layer consists of the hardware design written using HDLs like Verilog, and the cocotb layer consists of the testbench written using cocotb. These two layers interact with each other, simulating the DUT according to the verification logic. A conventional verification environment for hardware, written using cocotb, consists only of these two layers. The RL layer has been added on top of this to interact directly with the cocotb layer and indirectly with the hardware layer (that is, the DUT), completing the RL feedback loop. A more detailed block diagram representing what happens within a \textit{timestep}, along with which layer each block belongs to, is shown in Figure \ref{RL_verif_without_notation}.

As discussed, one of the objectives of the framework was to achieve modularity between the components. We share some of the technical details relevant to achieve this modularity. \\

\noindent \textbf{Signal exchange} The RL layer and the cocotb layer run in separate processes and interact with each other with the help of a \textit{pipe} object native to the multiprocessing module of Python. Through this pipe, the RL agent sends the action for the next step, as dictated by the policy it follows. The cocotb layer receives this action through the pipe and processes it to generate the exact input signal sequence that must be driven to the DUT as part of the step. While the DUT gets simulated with that input sequence, the elements of interest within the DUT, like registers, buffers, etc., are monitored using cocotb co-routines. Once the DUT simulation for that step completes, the information required for the Markov state and reward computation is communicated back to the RL layer via the pipe. The RL agent receives this information and uses it for training. \\

\noindent \textbf{Actions} The action space offered by VeRLPy consists of knobs, the values of which parameterize the input stimulus to the DUT. Each knob value is a real number lying in some user-specified interval and can collectively be used to specify the input signal sequence to the DUT for that timestep. The user can choose the number of such knobs required in the action space and specify how the input sequence to the DUT is generated from these knob values. Finite sets of discrete values can also be included as part of the action space. By default, VeRLPy assumes a single step, single Markov state environment as recommended in \cite{better}. However, VeRLPy also supports multi-step MDPs which require more careful integration between the RL and cocotb components. \\

\noindent \textbf{Rewards} A crucial design choice during the verification is to choose the reward signals. We define rewards as functions on top of functional events of interest tracked as part of the coverage on the DUT.  Let $e_i$ denote the $(i+1)$-th functional event tracked. $e_i$ could for instance be a certain register getting reset. The reward $R_t$ in each timestep $t$ is calculated based on user-specified multipliers for each such functional event tracked as part of coverage. The reward is formulated as, \begin{equation} \label{reward}
    R_t = \sum_{i=0}^{i=N-1}n_i \times m_i
\end{equation} where $N$ denotes the number of events being monitored in the step, $n_i$ denotes the number of times $e_i$ occurred in the step and $m_i$ denotes the user-specified multiplier that decides the contribution of $e_i$ to the reward. Thus, if there are $N = 3$ events being tracked as part of coverage and $m_0 = m_1 = 0, m_2 = 1$, the agent prefers actions which lead to higher coverage of event $e_2$ based on the reward function from equation \ref{reward}. This results in higher confidence in the verification of the logic associated with event $e_2$ and can lead to better chances of hitting corner cases that only occur a fraction of times when $e_2$ is hit.

The objective of the agent in a single-step RL setting is to increase the functional coverage of certain events of interest in this manner. These events could be chosen as those that occur less frequently in the DUT during simulation under a conventional random verification run. However, if $e_2$ corresponds to a coverage hole that very rarely occurs during exploration, the rewards will be too sparse for learning to happen in a reasonable amount of time. Tackling this challenge requires a multi-step MDP and RL algorithms for sparse reward environments.

\subsection{Control Flow}

\begin{figure*}[htbp]
\centerline{\includegraphics[width=10cm]{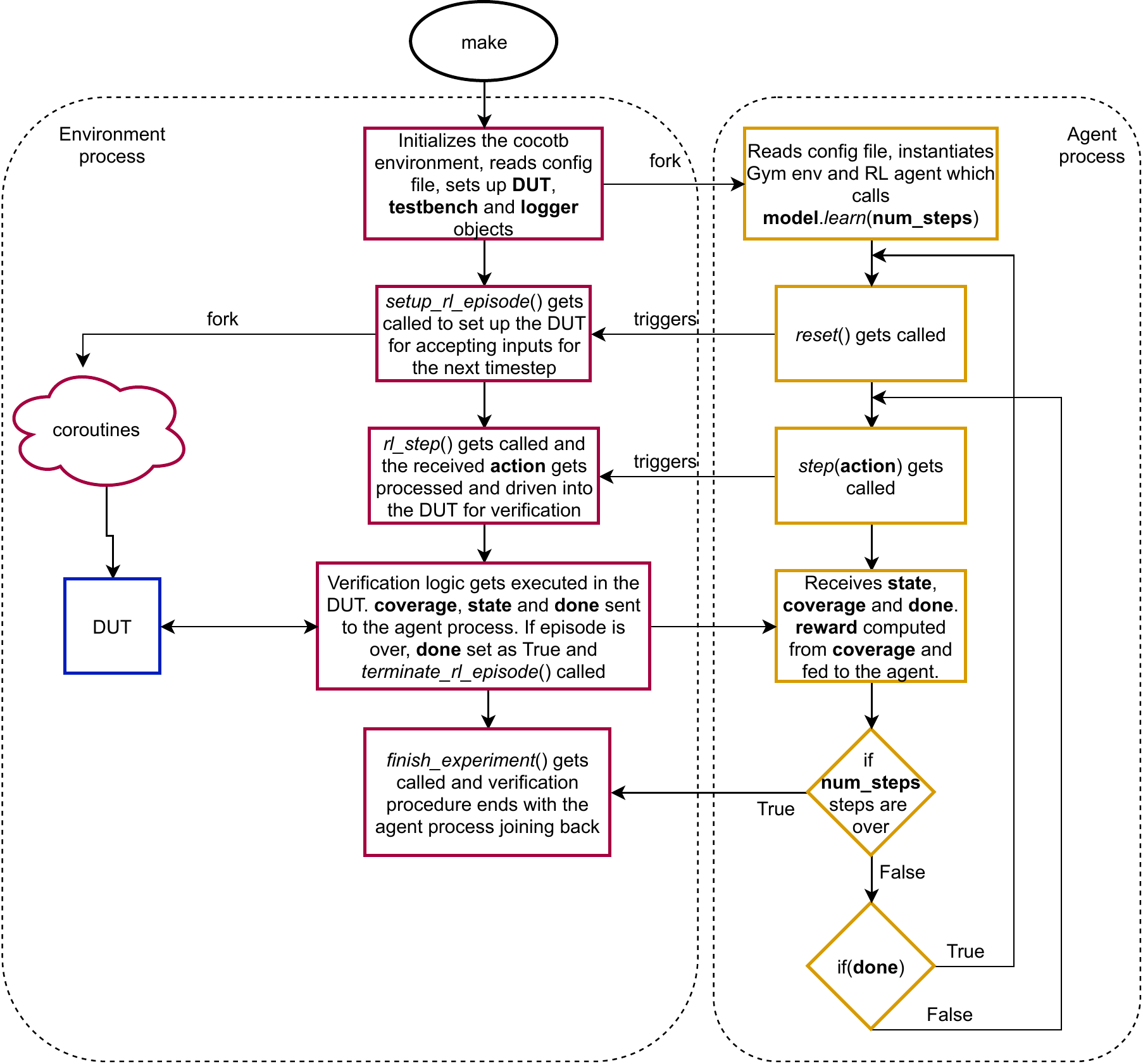}}
\caption{Control flow of VeRLPy. The two blocks shows the two processes - cocoTB process (left) and RL process (right) in the framework.}
\label{arch}
\end{figure*}

The verification logic defined in the cocotb layer are triggered appropriately when the \texttt{reset()} and the \texttt{step()} functions of the Gym environment get called. Abstract methods that can be overridden by the user are provided in VeRLPy for implementing the verification logic. These methods get called in the simulation loop in a specific sequence based on triggers from the RL agent and the RL environment. The cocotb layer architecture is designed with abstract methods to have VeRLPy usable for building verification setups for any digital design without having to deal with the details of interfacing between the RL and cocotb layers. Figure \ref{arch} shows the key abstract methods, their respective triggers, and the overall control flow of VeRLPy. 

The RL process starts running after the initialization of the DUT, the Gym environment, and the RL agent is done. Each RL episode starts with a call to the \texttt{reset()} function of the Gym environment. This triggers the cocotb logic to get the DUT ready for accepting inputs for the next timestep. Subsequently, the \texttt{step()} function of the Gym environment gets called one or more times depending on if the MDP is single-step or not. Each call of the \texttt{step()} function results in the action chosen by the RL agent being sent to the cocotb environment for getting processed and driven into the DUT for verification. The consequences of these inputs in the DUT are tracked using coroutines and used to compute the resulting Markov state changes and coverage-based rewards. Once these are sent back to the RL agent, the learning occurs, and its policy gets updated. Once all the timesteps of the episode are completed, the termination logic of the cocotb environment gets asserted and the next episode runs.

In addition to supporting the core RL functionality, VeRLPy also has auxiliary components for supporting an end-to-end process. A configuration file is defined to serve as the source for the RL action space, hyperparameters, reward function, and simulation parameters. There exists a mode where the verification can be performed without the RL feedback, for benchmarking against the traditional random baseline and easy debugging. Logging and visualization is also included for convenience. The logs contain the essential information from both the RL and cocotb layers like coverage, actions chosen, and rewards given but can be extended to include other design-specific information as well. The visualization scripts post-process the generated log files for creating the plots with all this data. 

\section{Experimental Results}
\label{sec:experiments}

In this section, we demonstrate the use of VeRLPy to verify two designs that are part of larger open source projects that we work on. 
We describe below the designs, and how VeRLPy is used. 
We also evaluate the use of VeRLPy by comparing the inputs that is driven to DUT and the occurence of required events achieved with and without RL when run for 1,000 iterations each. 
In both cases, we observe that the use of RL, as enabled by VeRLPy, significantly increases the occurence of events of interest in comparison to randomly chosen inputs. 

\subsection{RLE Compressor}
The first design we study is a Run Length Encoding (RLE) compressor which is used to store sparse matrices effectively.
This is a particularly important module in a deep learning accelerator \cite{shaktimaan} because it reduces the IO bandwidth limitation while reading sparse matrices from the DRAM onto the accelerator. 
The compressor distinguishes each element in the matrix as zero or non-zero. Non-zero values are stored in the compressed word vector and the continuous zeros in the matrix are stored as zero counts in the compressed zero count values vector. 
In addition to these two vectors, the compressor also holds the following registers for control flow. Word counter register to hold the count of valid words in the compressed word vector, zero counter register to hold the count of valid zero counts in the compressed zero count vector, counter register which counts the number of consecutive zero in the matrix whose output is stored in compressed zero count vector and next count register which says whether partial zero count value is stored in the compressed zero count vector.

The verification of a design such as the RLE compressor is challenging because it depends on randomly generating input sequences corresponding to different matrices. 
Randomly generated matrices do not efficiently capture sparsity patterns, which affect the RLE compressor's execution.
In particular, the execution of the DUT depends on the patterns of consecutive zeros in the input sequence. 
We model this input generation with an RL agent by parameterizing the probability of each element of the sequence being zero as the action space of the agent. 
In addition to this, the design configuration parameter \texttt{count\_width} which decides when the count overflows, as well as the length of the sequence are added as actions for the RL agent. 
The action space thus becomes the cross product of three sets of independent actions given as $[0, 1] \times \{1, 2, ..., 8\} \times \{100, 200, ..., 1000\}$. For example, if the action that gets chosen by the agent is (0.4, 6, 300), then a sequence of 300 natural numbers will be generated, where each number will be 0 with a probability of 0.4 and the \texttt{count\_width} parameter would be set as 6.

The next design choice is the reward signal for the RL agent. 
The reward signal is design specific and captures the designer's intent of the events of interest.
For the RLE compressor, we chose to track the following 4 events: compressed word vector counter reaching its maximum threshold (\texttt{word counter register} == 16), 
compressed zero count vector counter reaching its maximum threshold (\texttt{zero counter register} == 64), continuous count stored in the compressed zero count vector (\texttt{counter register} == $2^{\texttt{count\_width} - 2}$) and partial counts stored in the compressed zero count vector (\texttt{next count register} != 0).

We compare the use of RL in VeRLPy against randomized testing. 
We specifically target the event $e_3$ which corresponds to storing partial zero counts in the compressed zero vector. 
Given that the count vector is 64-bit long, the partial counts will be stored only when \texttt{count\_width} is not divisors of 64 that is, 3, 5, 6, and 7.
To specifically trigger $e_3$ in the DUT, we set the reward function for the RL agent to be positive when $e_3$ is triggered and be 0 otherwise. That is, $ m_0 = m_1 = m_2 = 0, m_3 = 1 $. 
1000 iterations of verification runs were done with and without RL feedback for comparison.

\begin{figure*}[htbp]
\centerline{
    \subfigure[\label{rle_comp_a}]{\includegraphics[width=0.33\hsize]{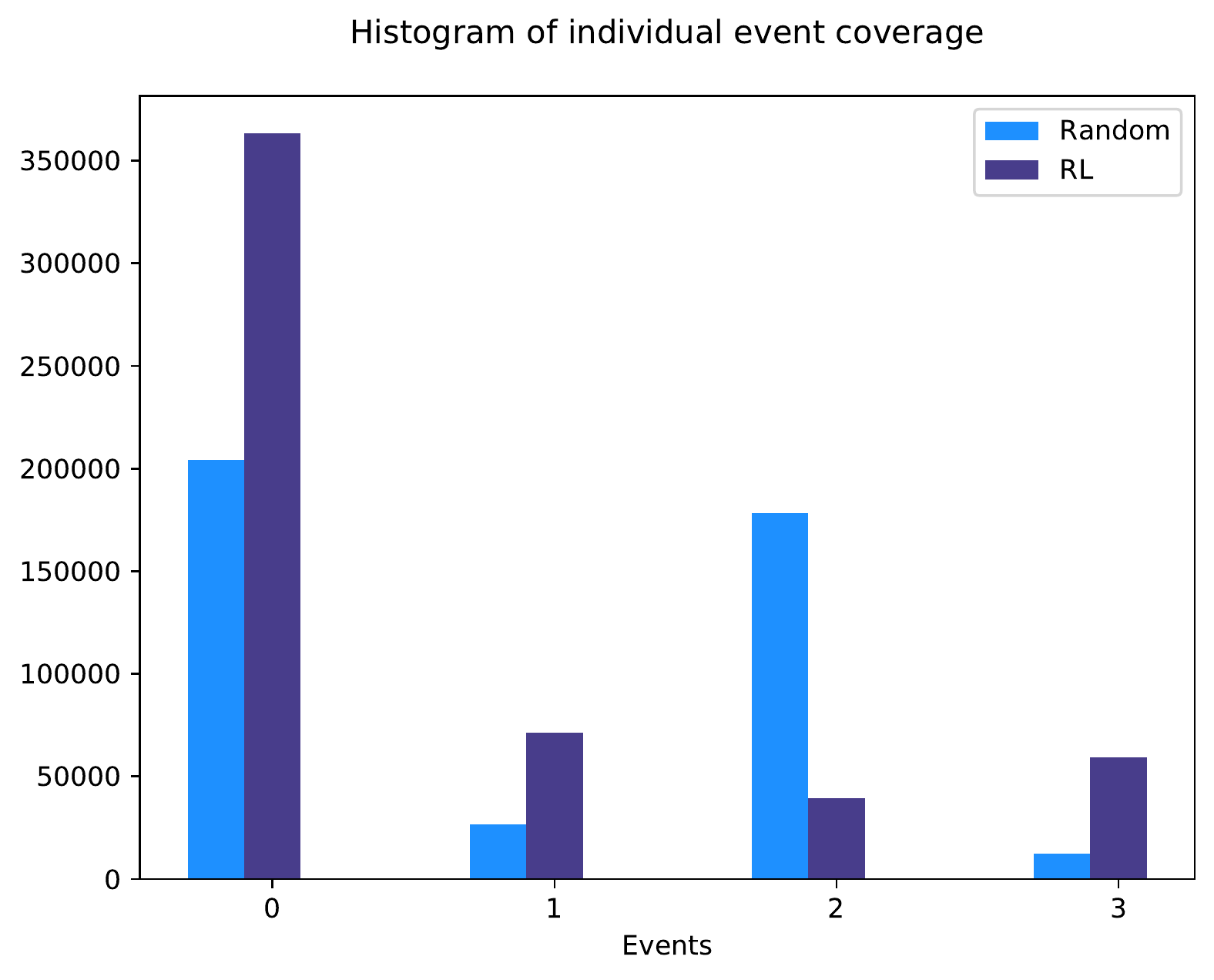}}
    \subfigure[ \label{rle_comp_b}]{\includegraphics[width=0.33\hsize]{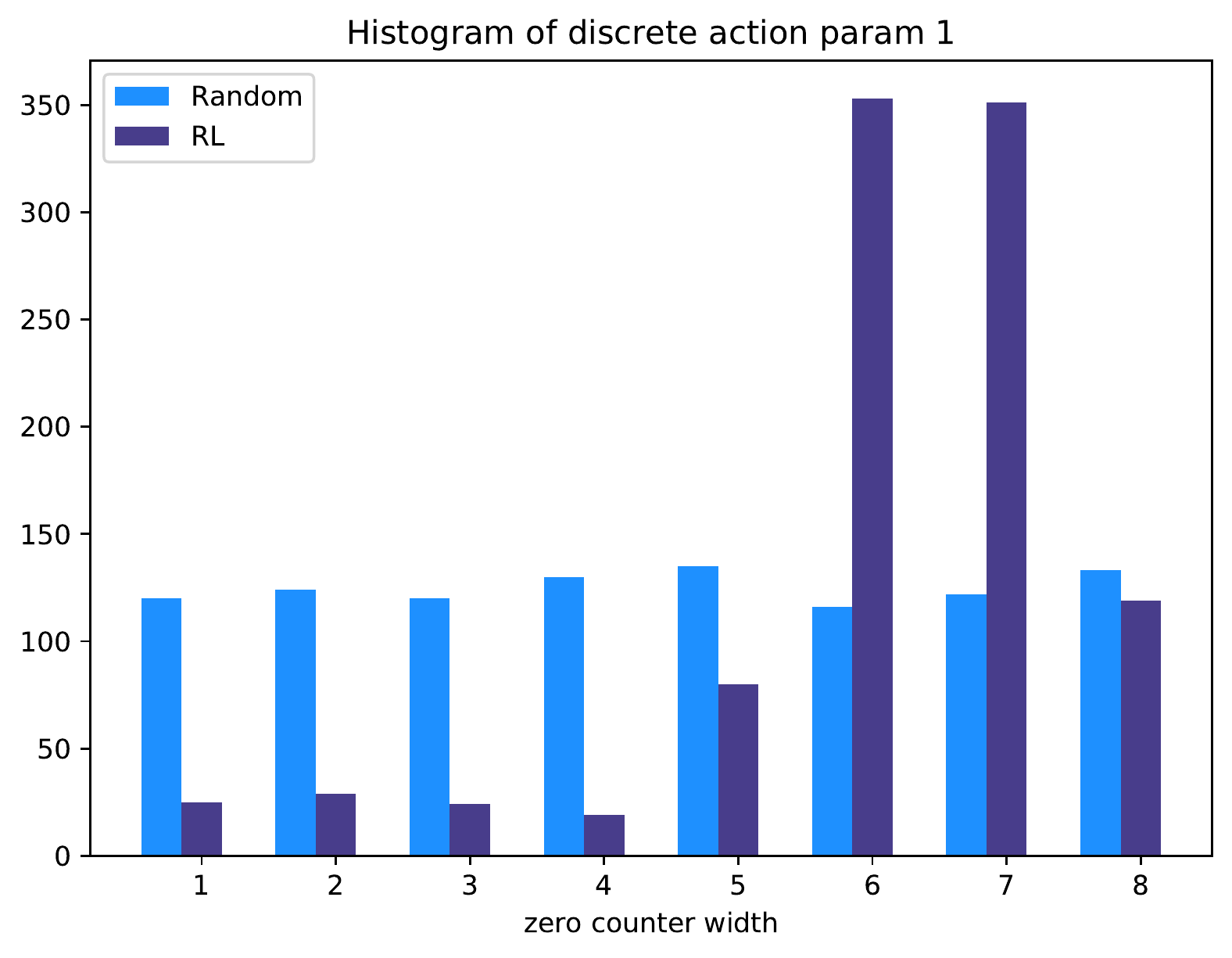}}
    \subfigure[ \label{rle_comp_c}]{\includegraphics[width=0.33\hsize]{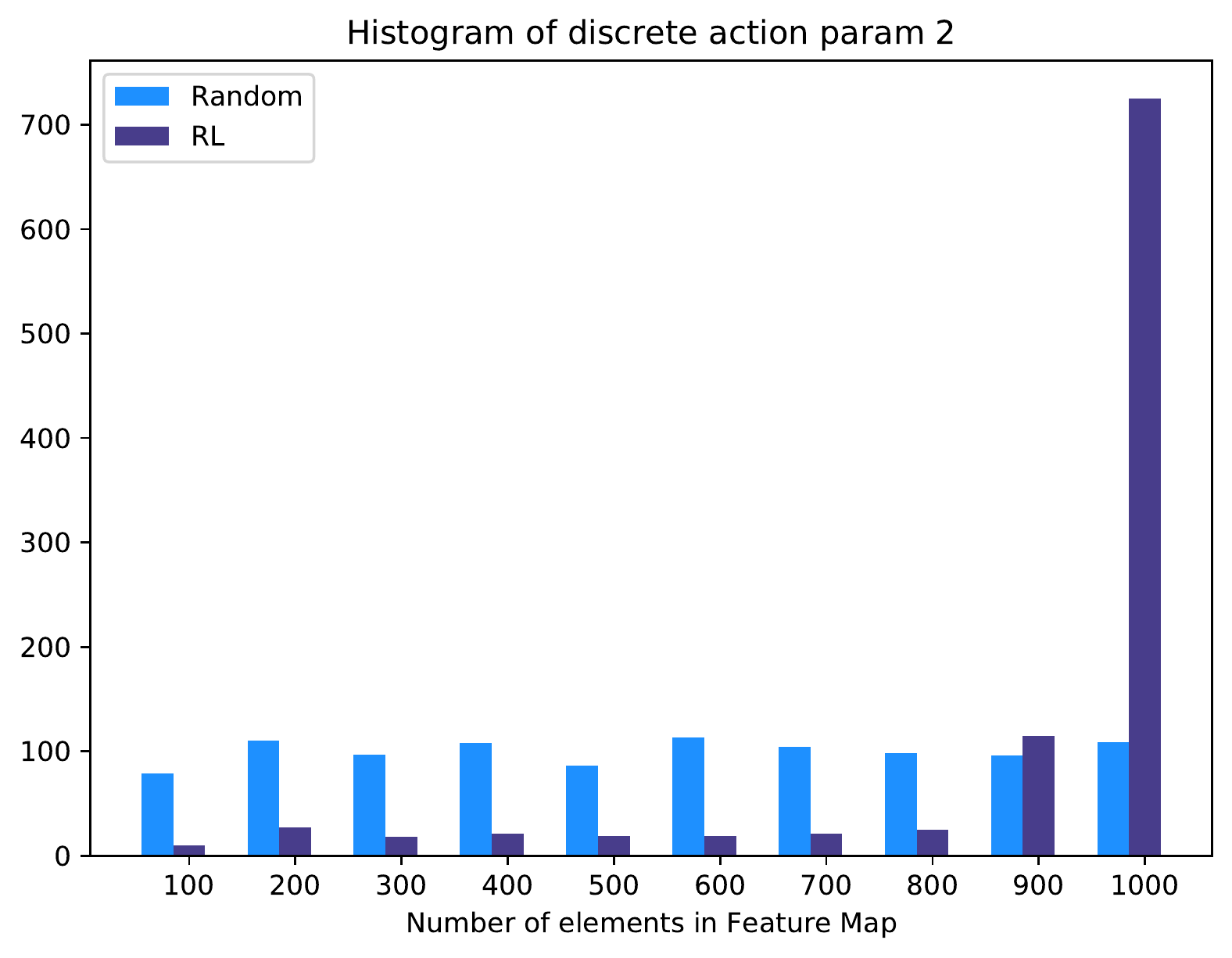}}
}
\centerline{
    \subfigure[ \label{axi4_a}]{\includegraphics[width=0.33\hsize]{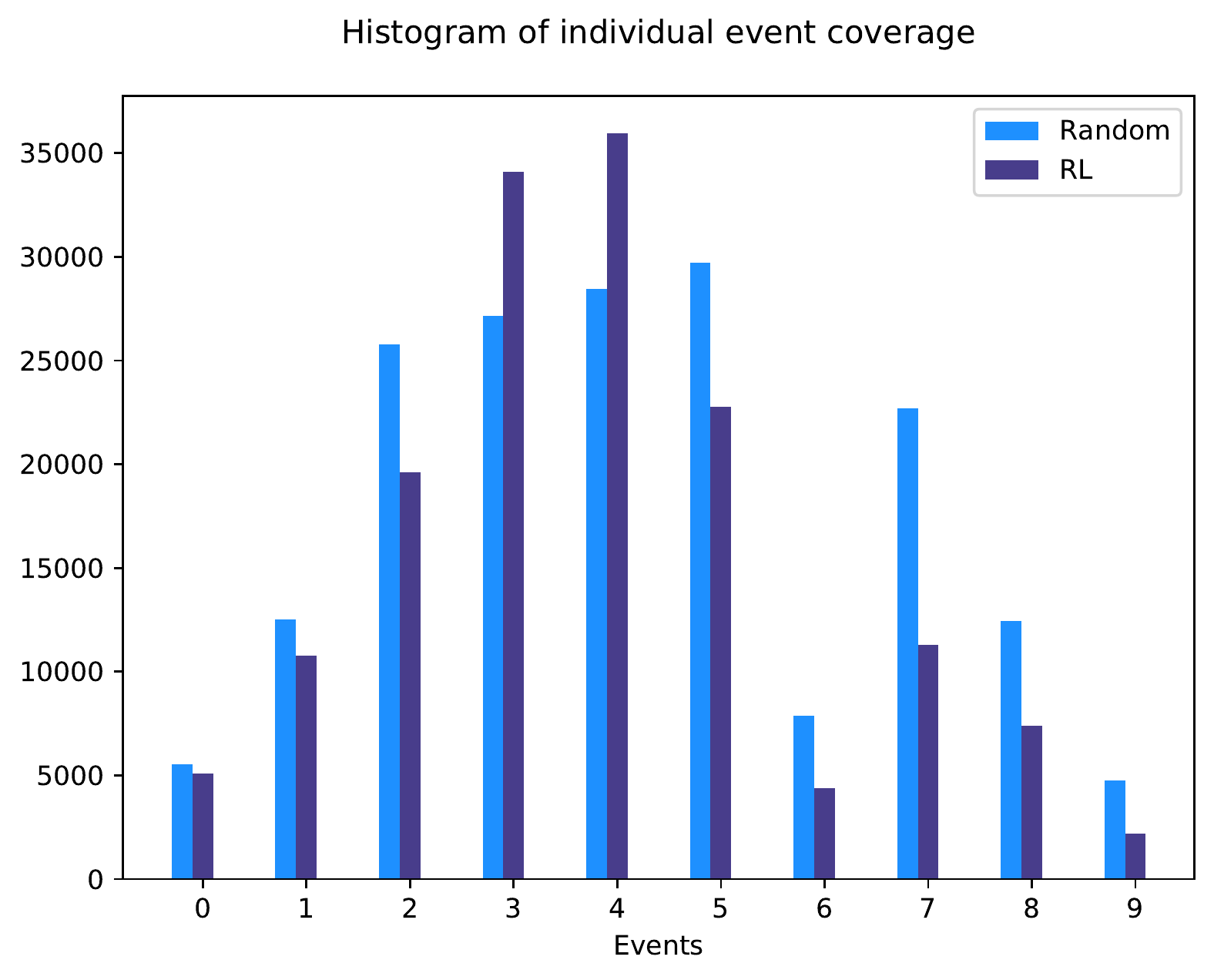}}
    \subfigure[ \label{axi4_b}]{\includegraphics[width=0.33\hsize]{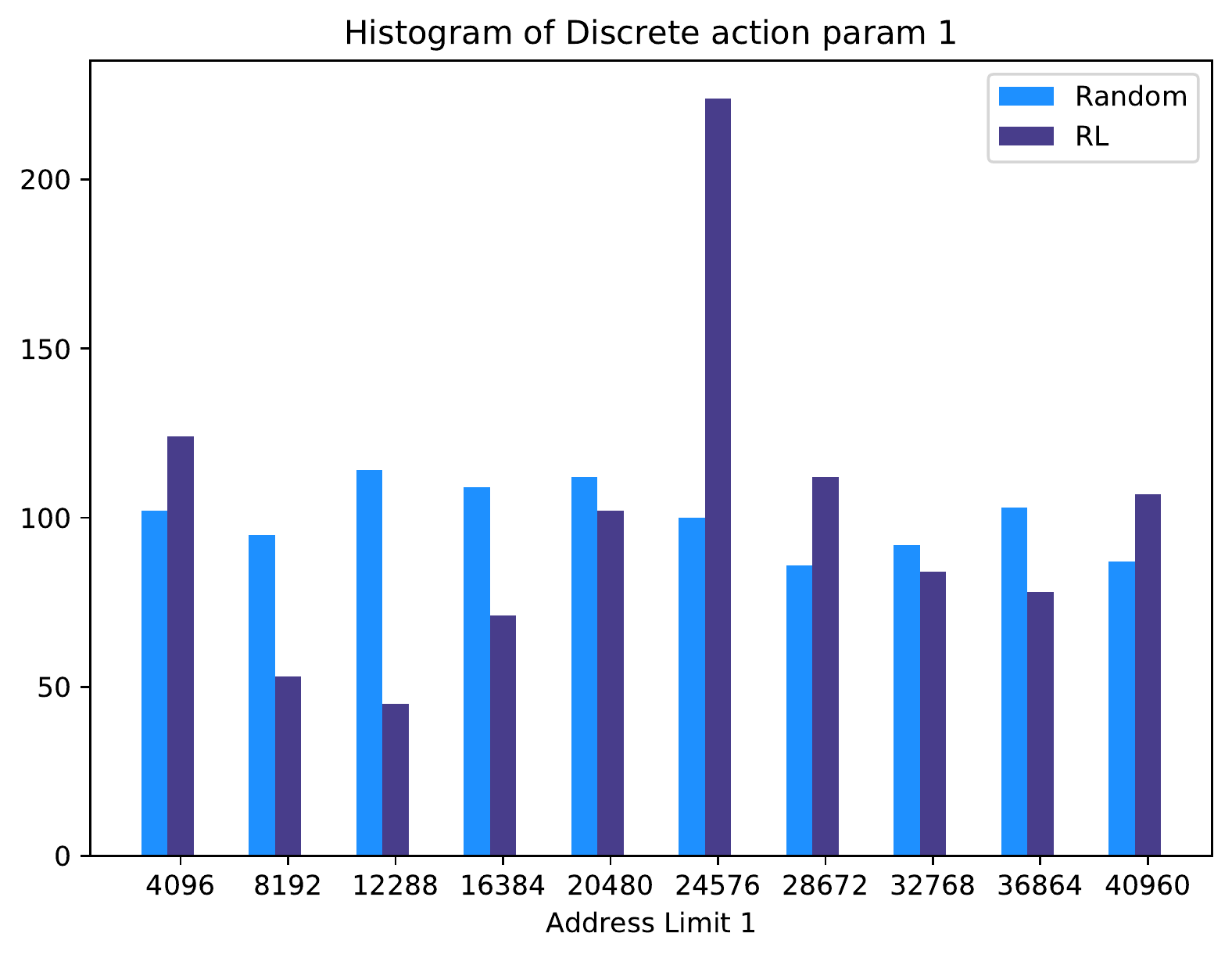}}
    \subfigure[ \label{axi4_c}]{\includegraphics[width=0.33\hsize]{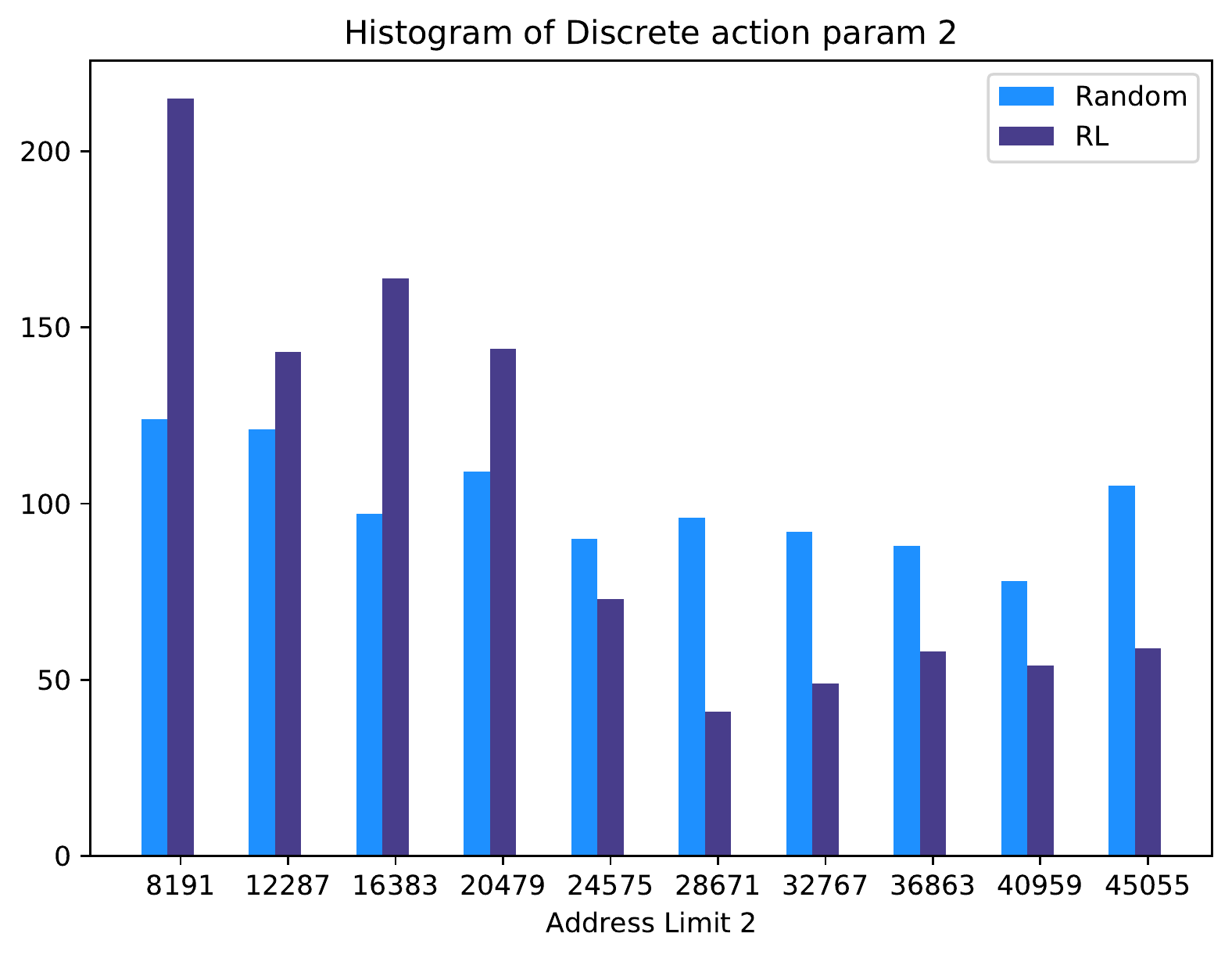}}
}
\caption{Comparison of coverage and actions chosen during the 1000 iterations. RLE Compressor - Figure \ref{rle_comp_a} shows the event coverage that is tracked when event $e_3$ is rewarded. Figure \ref{rle_comp_b} shows the histogram of \texttt{count\_width} values suggested by the RL agent. Figure \ref{rle_comp_c} shows the histogram of sequence length suggested by the RL agent. AXI Crossbar - Figure \ref{axi4_a} shows the event coverage that is tracked when event $e_4$ is rewarded. Figure \ref{axi4_b} and Figure \ref{axi4_c} shows the histograms of the choices for the lower and upper limit of the address ranges}
\label{}
\Description[RL improved the event coverage and suggested actions accordingly]{The histogram of event coverage shows that the occurrence of the e$3$ has increased and the RL has suggested appropriate count widths of 6 \& 7 and feature map length}
\end{figure*}

We plot the histogram of the different recorded events in both cases in Figure \ref{rle_comp_a}. 
As can be seen, the event $e_3$ occurred significantly more number of times with RL feedback than without (61,018 vs 12,290). 
This is particularly important because event $e_3$ is relatively rare in comparison with other events. 
So, adding the RL feedback without any other change is effective in adapting the input sequences to trigger the relatively rare event. 
It is also instructive to see the actions of the RL agent that cause the rare events. 
We plot the histograms of the actions suggested by RL in Figures \ref{rle_comp_b} and \ref{rle_comp_c}. 
Clearly, the RL agent has prioritised 6 and 7 as values of \texttt{count\_width} in comparison to the random algorithm. 
As discussed, since 6 and 7 are not divisors of 64, they cause the event $e_3$.
Notice that this specific insight on which values of the parameter lead to the event under consideration is learnt by the agent without any explicit input from the designer. This demonstrates that VeRLPy can be used to accelerate the coverage of rare events in the design.

\subsection{AXI Crossbar}

The AXI crossbar is a popular interconnect which transfers requests from masters to slaves and transfers responses from slaves to masters. 
It is a component that we developed in our effort to build open-source processors using the RISC-V instruction set architecture \cite{shakti}. 
The verification of the crossbar is challenging because the contention generated and the execution of the hardware depends on generating carefully timed sequences of requests and responses. 

We choose a particular example, where the crossbar has two masters and ten slaves. All the masters and the slaves have FIFOs to hold more than one request and response and each FIFO has \texttt{not\_empty} and \texttt{not\_full} registers which control the enqueueing and dequeueing of data. If \texttt{not\_full} register is False, enqueue into the FIFO is not allowed and if \texttt{not\_empty} register is False, dequeue from the FIFO is not allowed. 

The verification involved choosing a subset of the slaves and issuing requests from the masters only to those slaves. The RL action space was the lower and upper limits of the addresses of the slaves chosen. Thus, the action space was $\{$$ \texttt{lower\_slave\_addr}_0,$ $\texttt{lower\_slave\_addr}_1,$ $...,$ $\texttt{lower\_slave\_addr}_9 \}$  $\times$  $\{ \texttt{upper\_slave\_addr}_0,$ $\texttt{upper\_slave\_addr}_1,$ $...,$ $\texttt{upper\_slave\_addr}_9 \}$ where $\texttt{lower\_slave\_addr}_i$ and $\texttt{upper\_slave\_addr}_i$ correspond to minimum and maximum addresses taken by slave $i$ respectively. RL agent suggests one of the lower addresses and one of the upper addresses. The AXI request is generated between the minimum and the maximum of addresses suggested by RL.   

Given this requirement, we set up an RL agent which rewards the filling up of a slave FIFO buffer. 
More specifically we define event $e_i$ as denoting the FIFO full condition for slave $i$.
We run the RL agent specifically to reward $e_4$, i.e., to generate transfers and requests to overflow the FIFO of the 5th slave. 
In the absence of domain expertise, such a guided verification is not possible without the use of RL. 

We run the experiment for 1,000 iterations with and without the RL feedback. The results are plotted in Figure \ref{axi4_a} and show that the number of times event $e_4$ occurs is significantly higher with RL (49772 times with RL vs 25609 times without RL). 
Figures \ref{axi4_b} and \ref{axi4_c} also make it clear how the RL agent achieves this. It specifically chooses the address range (both min and max) to maximize the occurrence of event $e_4$.
Thus, we find that the RL agent learns the address decoding logic in the crossbar and generates inputs on address ranges such that the required slave achieves FIFO full condition more frequently. 
Again, this is done without any explicit cue from the designer. 

In summary, through the two real-world designs that we are working on for open-source projects, we showed how the RL agent can be used to identify inputs that guide the DUT to design-specific events. 
Without designer intuition, such guided verification cannot be performed without RL. 
This illustrates the possibilities for more cost-effective and efficient verification with VeRLPy.

\section{Conclusion and Future Directions}
\label{sec:conclusion}

VeRLPy is the first open-source end-to-end application framework to verify digital hardware designs with the aid of Reinforcement Learning. It is available at \cite{VeRLPy}. It is developed with the goals of modularity between the RL, verification, and hardware design layers, and extensibility given that it is a fully Python-based framework. With the increasing demand for more powerful and application-specific hardware in the post-Moore era, VeRLPy intends to serve the role of making the hardware development pipeline cost-effective and efficient.

VeRLPy also serves as a starting point for integrating more sophisticated RL ideas into the hardware verification process. One of the promising directions for further work would be to support goal-based environments so that approaches that build on Hindsight Experience Replay \cite{andrychowicz2018hindsight, lee2021deep} can be applied to hardware verification. Such methods might prove useful in the context of discovering bugs and coverage holes in the design. In addition to this, since the scale of the MDP could increase for more complex designs, intrinsic reward signals like Curiosity \cite{pathak2017curiositydriven} to tackle the large state space, and methods to efficiently act in large action spaces like Wolpertinger policy \cite{dulacarnold2016deep} are worth pursuing using VeRLPy.



\bibliographystyle{ACM-Reference-Format}
\bibliography{VeRLPy}

\end{document}